\documentclass[reprint, amsmath, amssymb, aps, nofootinbib, twocolumn,superscriptaddress]{revtex4}
\usepackage{graphicx}% Include figure files
\usepackage{dcolumn}% Align table columns on decimal point
\usepackage{bm}% bold math
\usepackage{comment}
\usepackage{fontawesome}
\usepackage{aas_macros}
\usepackage{tabularx}
\usepackage{multirow}
\usepackage{xcolor}
\usepackage[colorlinks,urlcolor=blue,citecolor=blue,linkcolor=purple]{hyperref}

\newcolumntype{P}[1]{>{\centering\arraybackslash}p{#1}}

\newcommand{\orcid}[1]{$\,$\href{https://orcid.org/#1}{\includegraphics[width=10pt]{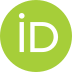}}}

\begin{document}

\title{Weighing the Milky Way and Andromeda with Artificial Intelligence}

\author{Pablo Villanueva-Domingo\orcid{0000-0002-0936-4279}}
\email{pablo.villanueva.domingo@gmail.com}
\affiliation{Instituto de F\'isica Corpuscular (IFIC), CSIC-Universitat de Val\`encia, E-46980, Paterna, Spain}

\author{Francisco Villaescusa-Navarro\orcid{0000-0002-4816-0455}}
\email{fvillaescusa@flatironinstitute.org}
\affiliation{Center for Computational Astrophysics, Flatiron Institute, 162 5th Avenue, New York, NY, 10010, USA}
\affiliation{Department of Astrophysical Sciences, Princeton University, Peyton Hall, Princeton NJ 08544, USA}

\author{Shy Genel\orcid{0000-0002-3185-1540}}
\affiliation{Center for Computational Astrophysics, Flatiron Institute, 162 5th Avenue, New York, NY, 10010, USA}
\affiliation{Columbia Astrophysics Laboratory, Columbia University, New York, NY, 10027, USA}

\author{Daniel Angl\'es-Alc\'azar\orcid{0000-0001-5769-4945}}
\affiliation{Department of Physics, University of Connecticut, 196 Auditorium Road, U-3046, Storrs, CT 06269-3046, USA}
\affiliation{Center for Computational Astrophysics, Flatiron Institute, 162 5th Avenue, New York, NY, 10010, USA}

\author{Lars Hernquist}
\affiliation{Center for Astrophysics — Harvard \& Smithsonian, 60 Garden St, Cambridge, MA 02138, USA}

\author{Federico Marinacci}
\affiliation{Dipartimento di Fisica e Astronomia ‘Augusto Righi’ , Universit\`a di Bologna, via Gobetti 93/2, 40129, Bologna, Italy}

\author{David N. Spergel}
\affiliation{Center for Computational Astrophysics, Flatiron Institute, 162 5th Avenue, New York, NY, 10010, USA}
\affiliation{Department of Astrophysical Sciences, Princeton University, Peyton Hall, Princeton NJ 08544, USA}

\author{Mark Vogelsberger}
\affiliation{Kavli Institute for Astrophysics and Space Research, Department of Physics, MIT, Cambridge, MA 02139, USA}

\author{Desika Narayanan}
\affiliation{Department of Astronomy, University of Florida, Gainesville, FL, USA}
\affiliation{University of Florida Informatics Institute, 432 Newell Drive, CISE Bldg E251, Gainesville, FL, USA}

\begin{abstract}
We present new constraints on the masses of the halos hosting the Milky Way and Andromeda galaxies derived using graph neural networks. Our models, trained on thousands of state-of-the-art hydrodynamic simulations of the CAMELS project, only make use of the positions, velocities and stellar masses of the galaxies belonging to the halos, and are able to perform likelihood-free inference on halo masses while accounting for both cosmological and astrophysical uncertainties. Our constraints are in agreement with estimates from other traditional methods.
%\href{https://github.com/PabloVD/HaloGraphNet}{GitHub \faicon{github}}.
\end{abstract}

\maketitle

\section{Introduction}
\vspace{-0.25cm}

The $\Lambda$CDM model describes how non-linear gravitational evolution amplifies the small dark matter perturbations after inflation creating larger and deeper gravitational potential wells where gas can collapse, cool down, and form stars and galaxies. The observed rotation curves of galaxies indicates the existence of an additional mass beyond the one that can be accounted for from luminous matter, indicating that galaxies are embedded in larger objects called halos; this is in agreement with the results of hydrodynamic simulations. Determining the mass of the halo hosting our galaxy or that of our neighbors is important in order to perform an independent check of the $\Lambda$CDM model, and would provide us with a better understanding of the dark matter-baryon relation. Furthermore, knowing the total mass of our galaxy and its companions will help us to determine their fate \cite{Cautun_2019, Marel_2012}.

In \cite{GNN_CAMELS} we developed a new machine learning method that uses graph neural networks (GNNs) to perform likelihood-free inference of halo masses from the positions, velocities, and internal properties of the galaxies they host. Our model does not make any assumption about the state of the system, and since it was trained on galaxies from simulations of the Cosmology and Astrophysics with MachinE Learning Simulations (CAMELS) project \cite{villaescusanavarro2020camels}, automatically marginalizes over cosmological and astrophysical uncertainties. We also showed that the model predictions are robust, i.e. the outcome does not depend on the particular type of simulation used to train the GNN. In this work we use those networks to infer the masses of the halos hosting the Milky Way (MW) and Andromeda (M31) galaxies, being, to our knowledge, the first time that artificial intelligence is applied to this task.

\begin{table*}[th!]
	\begin{center}
	%\resizebox{\textwidth}{!}{
		%\begin{tabular}{|P{0.08\linewidth}|P{0.19\linewidth}|P{0.12\linewidth}|P{0.15\linewidth}|P{0.32\linewidth}|}
		\setlength{\tabcolsep}{12pt}    % row separation
	\renewcommand{\arraystretch}{1.2}   % column separation
		\begin{tabular}{|c|c|c|c|c|c|}
			\hline
			Halo & Galaxy & Position [kpc] & Velocity [km/s] & Stellar mass $[M_\odot]$ & References \\ 
			\hline\hline
			\multirow{3}{*}{\, MW \,} & MW & $(0,0,0)$ & $0.0$ & $5\times 10^{10}$ & \cite{2016ARAA..54..529B}  \\
			\cline{2-6}
			& LMC & $(-0.6, -41.3, -27.1)$ & $322.1$ & $2.7\times 10^9$ & \cite{2002AJ....124.2639V,  2013Natur.495...76P, 2015arXiv151103346B, 2021AA...649A...7G} \\
			\cline{2-6}
			& SMC & $(15.8, -37.3, -43.3)$ & $237.4$ & $3.1 \times 10^8$ & \cite{2014MNRAS.442.1663D, 2015arXiv151103346B, 2016ApJ...816...49S, 2021AA...649A...7G} \\
			\cline{2-6}
			%& Sgr dSph & $(17.5, 2.5, -6.5)$ & $317.6$ & $10^{8}$ & \cite{2020MNRAS.497.4162V} \\
			\hline\hline
			\multirow{4}{*}{\, M31 \,} & M31 & $(-386.0, 622.7, -287.3)$ & $0.0$ & $1.0\times 10^{11}$ & \cite{2012AJ....144....4M, 2015IAUS..311...82S, 2019ApJ...872...24V, 2012AA...546A...4T}  \\
			\cline{2-6}
			& M33 & $(-485.9, 500.6, -419.0)$ & $257.6$ & $2.9\times 10^9$ & \cite{2012AJ....144....4M, 2019ApJ...872...24V} \\
			\cline{2-6}
			& M110 & $(-401.4, 661.0, -295.6)$ & $116.1$* & $3.3 \times 10^8$ & \cite{2012AJ....144....4M} \\
			\cline{2-6}
			& M32 & $(-395.5, 638.4, -300.6)$ & $38.1$* & $3.2\times 10^8$ & \cite{2012AJ....144....4M} \\
			%\cline{2-6}
			%& NGC 147 & $(-357.2, 608.8, -178.3)$ & 188.9 & $10^8$ & \cite{2020ApJ...901...43S, 2013ApJ...779..102K} \\
			%\cline{2-6}
			%& NGC 185 & $(-323.5, 529.0, -158.1)$ &  & $ \times 10^8$ & \cite{2020ApJ...901...43S} \\
			\hline
		\end{tabular}
		%}
	\end{center}
	\vspace{-0.5cm}
	\caption{We use graph neural networks (GNNs) to infer the masses of the halos hosting the Milky Way (MW) and the Andromeda (M31) galaxies. This table shows the galaxies taken into account and their properties for both cases. The input data comprises the position in Cartesian galactocentric coordinates, the modulus of the velocity in the central galaxy rest frame (either MW or M31), and the stellar mass of each galaxy. Velocities marked with an asterisk are estimated by multiplying their radial components by $\sqrt{3}$.}
	\label{table:galaxies}
\end{table*}

\vspace{-0.5cm}
\section{Model}
\vspace{-0.25cm}

Our model works as follows. Given a halo, we first select all the galaxies belonging to it. Next, we create a mathematical graph, where each of the halo galaxies (including both central and satellites) are considered as nodes that are connected by edges depending on their relative distances. The nodes can have several properties associated to them, such as positions, velocities, and stellar masses of the galaxies they represent.

GNNs are deep learning architectures appropriate to deal with irregular data, like the graphs constructed above \cite{2021arXiv210413478B, 2018arXiv180601261B, HamiltonBook}. They exhibit many advantages over traditional convolutional neural networks (CNNs), and given the sparsity, structure and irregularity of the data we are working with, GNNs are more suited for this problem. They employ the so-called message passing scheme, where the node features of neighbors are aggregated to infer global quantities of the graph. In that way, GNNs exploit neighborhood relations and locality. These models are by construction permutation invariant, and we also perform data augmentation to enforce the network to be rotationally invariant. The GNNs are trained to perform likelihood-free inference of the halo mass, in order to output the posterior mean and standard deviation without making any assumption on the form of the posterior distribution \cite{moment_networks}. We use $M_{200c}$, the mass within a virial radius with a density 200 times the critical density, as the definition of halo mass. We refer the reader to \cite{GNN_CAMELS} for further details on the architecture and training procedure.

We train the networks on halos/galaxies from the CAMELS simulations \citep{villaescusanavarro2020camels,2021arXiv210910915V}, containing two different simulation suites that were run with two distinct codes employing different subgrid physics models: 1) AREPO\footnote{\url{https://arepo-code.org/}} \citep{Weinberger:2019tbd} and the IllustrisTNG model \citep{2017MNRAS.465.3291W, 2018MNRAS.473.4077P, 2019ComAC...6....2N}, and 2) GIZMO\footnote{\url{http://www.tapir.caltech.edu/~phopkins/Site/GIZMO.html}} \citep{Hopkins:2014qka} and the SIMBA model \citep{Dave:2019yyq}. These models differ in their treatment of astrophysical processes such as supernova and active galactic nuclei (AGN) feedback or black hole growth. Each suite (IllustrisTNG and SIMBA) contains 1,000 simulations, each of them having different values of two cosmological parameters ($\Omega_{\rm m}$ and $\sigma_8$) and four astrophysical parameters controlling the efficiency of feedback from supernovae and AGNs.\footnote{This corresponds to the CAMELS LH set \cite{villaescusanavarro2020camels, GNN_CAMELS}.} The halos and subhalos have been identified with the \textsc{SUBFIND} algorithm \citep{Subfind}. In the simulations, galaxies are defined as subhalos that contain more than 10 star particles. In this paper we only consider galaxy properties that can be accessed via observations:\footnote{We note that the models in \cite{GNN_CAMELS} also include the stellar half-mass radius as a galactic feature. However, since it may be difficult to estimate it observationally for some satellites, we do not include it in our analysis. Excluding this feature does not dramatically decrease the accuracy, presenting mean relative errors (in the logarithm of the mass) between $\sim 1- 2$ \%.} 1) positions, 2) modulus of the velocity, and 3) stellar mass of each galaxy. Positions and velocities of galaxies are defined relative to the rest frame of the central galaxy, about its origin. In this work we have trained four different networks: 1) using IllustrisTNG galaxies including velocities, 2) using IllustrisTNG galaxies without velocities, 3) using SIMBA galaxies including velocities, and 4) using SIMBA galaxies without velocities. We note that since the networks are trained employing galaxies having different cosmological and astrophysical models, the network learns to marginalize over these. We emphasize that in this setup we are considering all CAMELS models equally likely. In other words, we are assuming a flat prior on the value of the cosmological and astrophysical parameters.

\begin{figure*}[th!]
\begin{center}
\includegraphics[width=0.49\linewidth]{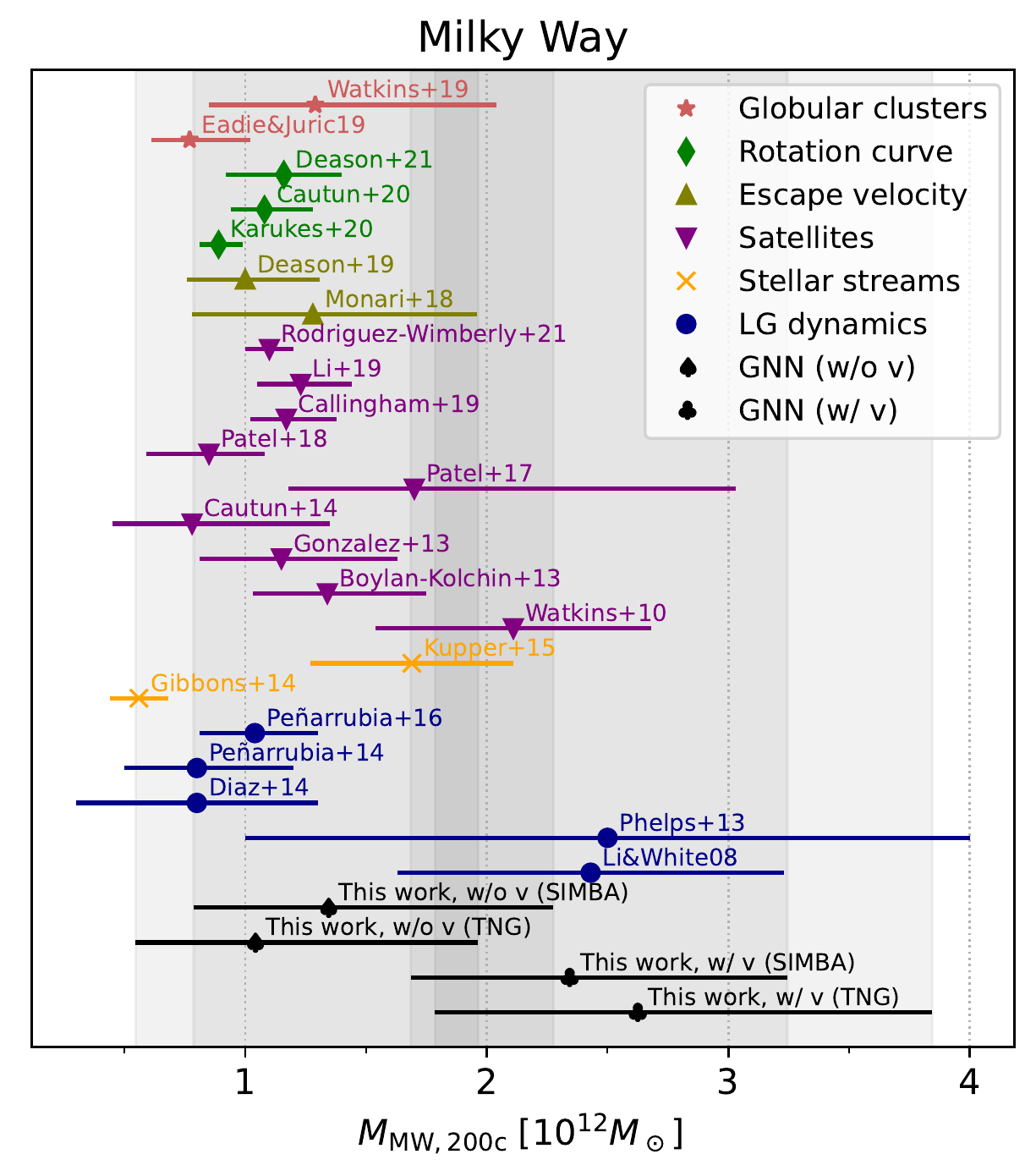}
\includegraphics[width=0.49\linewidth]{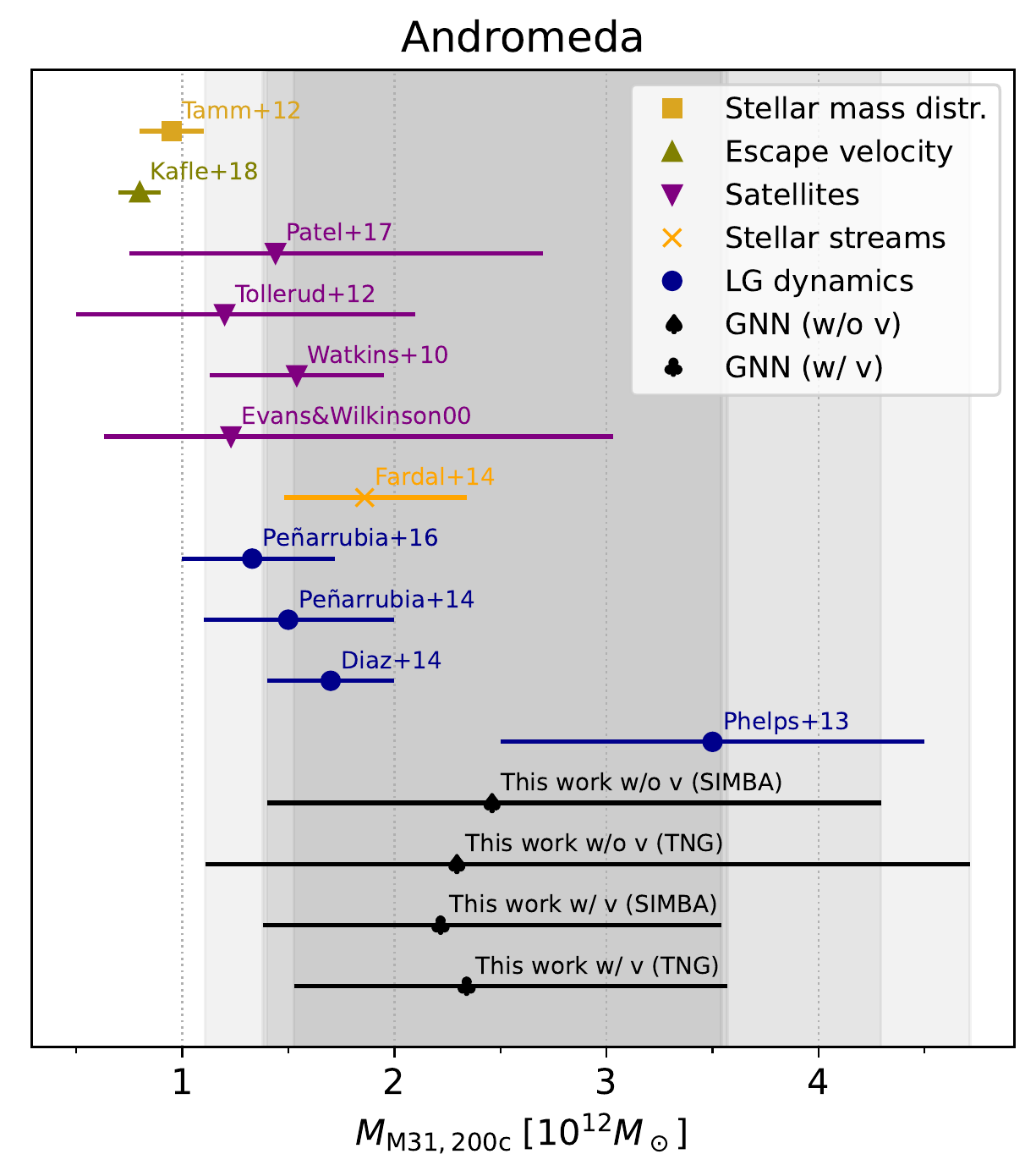}
\vspace{-0.4cm}
\caption{The colored points with error bars show constraints on the mass ($M_{200c}$) of the halo hosting the Milky Way (left) and the Andromeda (right) galaxies using different methods. The black points with error bars display instead the halo masses derived from graph neural network models that are trained from CAMELS simulations with and without velocities of the galaxies.}
\label{fig:massbounds}
\end{center}
\end{figure*}

\vspace{-0.5cm}
\section{Data}
\vspace{-0.25cm}

With the networks trained on simulated galaxies we apply them to real data. The input to the model are the positions, velocities, and stellar masses of the galaxies belonging to the halos hosting the MW and M31 galaxies. The list of known satellite galaxies of the MW has grown during the last years up to several dozens thanks to observations by DES \cite{Drlica-Wagner:2015ufc, 2018ApJS..239...18A} and SDSS \cite{2012ApJS..203...21A}. The PAndAS survey has also extended the number of M31 satellites above 30 \cite{2011ApJ...732...76R}.\footnote{See \cite{2012AJ....144....4M,2018MNRAS.479.2853N} for comprehensive compilations of the MW and M31 satellite galaxy data.} When building our graphs we only consider galaxies with stellar masses above $\sim 10^8~h^{-1}M_\odot$, since those are the galaxies present in the CAMELS simulations and therefore the ones used to train the GNNs.

The MW graph contains thus the Large and Small Magellanic Clouds (LMC and SMC), together with the Milky Way (MW) itself. The remaining dwarf satellites are much less massive \cite{Strigari:2008ib}. Stellar masses are extracted from \cite{2016ARAA..54..529B, 2015arXiv151103346B}, while velocities for the Magellanic Clouds are computed from radial velocities and proper motions from \cite{2002AJ....124.2639V, 2013Natur.495...76P, 2021AA...649A...7G, 2014MNRAS.442.1663D, 2016ApJ...816...49S}. Positions in the galactic coordinate frame have been transformed to cartesian galactocentric frame making use of astropy\footnote{\url{https://www.astropy.org}} \cite{astropy:2013, astropy:2018}.

The Andromeda graph contains M31, M33 (also known as the Triangulum galaxy), M32 (also known as Le Gentil), and M110 (also known as NGC 205). Note that the membership of M33 to the M31 halo is still under discussion, but the general consensus is that it is a satellite \cite{2012ApJ...753....8V, 2017MNRAS.468.3428P, 2020MNRAS.493.5636T}. We take the positions and stellar masses of Andromeda and its satellites from \cite{2012AJ....144....4M}. The galactocentric velocities for M31 and M33 are extracted from \cite{2019ApJ...872...24V}, using $\vec{v}_{\rm 3D, M31}=(34, -123, -19)$ km/s from their DR2+HST weighted average. Note however that there is not complete agreement about the tangential velocity of M31, which strongly relies on the specific approach followed \cite{2012ApJ...753....8V, 2016MNRAS.456.4432S, 2019ApJ...872...24V}. 3D velocities for M32 and M110 have not been derived in the literature, since their proper motions are difficult to measure, given their proximity to M31; only radial velocities have been estimated for these galaxies. Given that minor variations in the velocity of these small satellites are not expected to have a significant impact on the results, since only the modulus of the relative velocity is required rather than specific 3D values, and as long as other satellites also present large uncertainties in their velocities, we set their velocities to their radial component in the M31 frame multiplied by a factor of $\sqrt{3}$. We have verified that the precise value of the velocities of M32 and M110 does not have a major impact on the determination of the M31 mass. Table \ref{table:galaxies} presents a summary of the features and sources considered for each halo.

\vspace{-0.5cm}
\section{Results}
\vspace{-0.25cm}

Several distinct methods have been used to infer the mass of the MW halo: 1) from Local Group dynamics \cite{Li:2007eg, 2013ApJ...775..102P, 2014MNRAS.443.1688D, 2014MNRAS.443.2204P, 2016MNRAS.456L..54P}, 2) modelling stellar tidal streams \cite{2014MNRAS.445.3788G, 2015ApJ...803...80K}, 3) rotation curve data \cite{2015PASJ...67...75S, 2019JCAP...10..037D, Karukes:2019jwa, 2020MNRAS.494.4291C}, 4) the escape speed of the galaxy \cite{2018A&A...616L...9M, 2019MNRAS.485.3514D, 2021MNRAS.501.5964D}, 5) properties and dynamics of satellite galaxies \cite{2013ApJ...768..140B, 2013ApJ...770...96G, Cautun:2014dda, 2017MNRAS.468.3428P, 2018ApJ...857...78P, 2019MNRAS.484.5453C, 2020ApJ...894...10L, 2020MNRAS.494.5178F, 2021arXiv210900633R} and 6) halo globular cluster kinematics \cite{2019ApJ...875..159E, 2019ApJ...873..118W}. See, e.g., \cite{Wang:2019ubx, Karukes:2019jwa}, for recent comprehensive compilations of estimates of the MW mass using different approaches. Some of these methods have also been devoted to estimate the mass of the Andromeda halo, such as 1) from the Local Group dynamics \cite{2013ApJ...775..102P, 2014MNRAS.443.1688D, 2014MNRAS.443.2204P, 2016MNRAS.456L..54P}, 2) stellar streams \cite{2013MNRAS.434.2779F}, 3) rotation curve data \cite{2015PASJ...67...75S}, 4) the escape velocity of planetary nebulae \cite{2018MNRAS.475.4043K}, 5) satellite phenomenology \cite{2000MNRAS.316..929E, 2000ApJ...537L..91C, 2010MNRAS.406..264W, 2012ApJ...752...45T, 2017MNRAS.468.3428P}, 6) globular clusters \cite{2014MNRAS.442.2929V} and 7) the distribution of its stellar mass \cite{2012A&A...546A...4T}. Other works have inferred instead the sum of the MW and M31 total masses \cite{2012ApJ...753....8V, Gonzalez:2013pqa, 2017JCAP...12..034M, 2017MNRAS.465.4886C, 2021PhRvD.103b3009L}. The colored points in Fig. \ref{fig:massbounds} show the constraints derived for the MW (left) and M31 (right) halos.

The black points of Fig. \ref{fig:massbounds} show instead the bounds derived for the MW and M31 halo masses from the four GNNs mentioned above using the two training simulation suites and whether they incorporate peculiar velocities as features. Table \ref{table:masspredictions} summarizes these results. As can be seen, the predictions from models employing only positions and stellar mass (shown as black $\spadesuit$ points) are generally in good agreement with most of the previous literature, within the confidence limits. The estimated uncertainties are slightly larger than those from other traditional techniques, mainly due to the few features considered and given that the GNNs have been trained on simulations varying cosmological and astrophysical parameters. Predictions of the M31 halo mass taking into account also velocities (shown as black $\clubsuit$ points in the right panel of Fig. \ref{fig:massbounds}) are also consistent with standard estimations, but show smaller uncertainties, since the model is more precise when employing an additional feature.

\begin{figure}[t!]
\begin{center}
\includegraphics[width=0.99\linewidth]{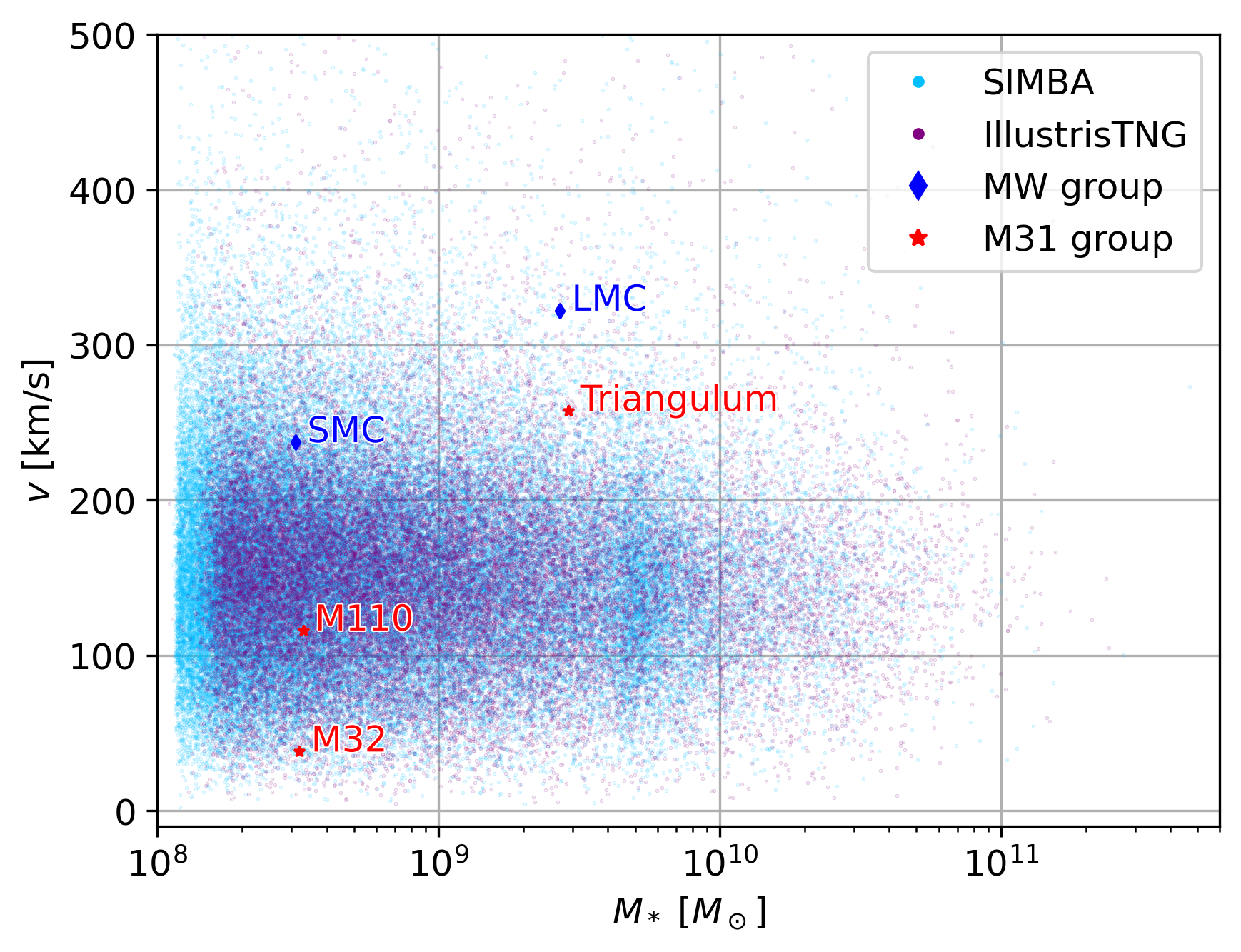}
\vspace{-0.8cm}
\caption{Velocity $v$ versus stellar mass $M_*$ for satellite galaxies within MW-like halos, with total mass between $0.8 \times 10^{12}$ and $1.5 \times 10^{12} M_\odot$, in the CAMELS simulations. Scatter points are extracted from the IllustrisTNG and SIMBA simulations suites. Satellite galaxies from the MW and M31 systems are also shown.}
\label{fig:vel_Mstar}
\end{center}
\end{figure}

However, we find that the predictions for the MW employing a GNN where galaxy velocities are included (shown as black $\clubsuit$ points in the left panel of Fig. \ref{fig:massbounds}) lead to a slightly larger halo mass. While these networks are more accurate, these predictions are in worse agreement with most of the literature estimates (although still consistent with some estimates from satellite and Local Group dynamics). We believe this result can be attributed to the fact that the velocity of the LMC is relatively large. To see this, we show in Fig. \ref{fig:vel_Mstar} the velocities versus stellar masses of galaxies within MW-like halos (with total masses between $0.8 \times 10^{12}$ and $1.5 \times 10^{12} M_\odot$) from both CAMELS suites: IllustrisTNG and SIMBA. Local Group satellite galaxies considered are also shown. It can be noted that the LMC presents a relatively large velocity compared to most CAMELS satellites of the same stellar mass. We believe that this large value of the most massive satellite would lead the GNN to predict a larger value of the halo mass. This fact is also consistent with other works in the literature employing velocities to infer the MW mass. For instance, the LMC velocity suggests a MW mass estimate larger than usual \cite{2011MNRAS.414.1560B, 2017MNRAS.468.3428P}. Moreover, the velocities of LMC and SMC-like satellites are found to be larger when the Local Group environment is taken into account \cite{2013ApJ...770...96G}. This goes along with the fact pointed out in previous works that the MW may behave as a somewhat unusual galaxy for some aspects, e.g., regarding its satellite population \cite{2017ApJ...847....4G} or its luminosity and rotational velocity relation \cite{1996ApJ...473..687M, 2016ApJ...833..220L}.

It is worth examining the robustness of the GNN predictions. Given that the networks have been trained applying random rotations to the galaxies in the halos, it is expected that the inference would be rotationally invariant. By performing 50 arbitrary rotations of the MW and M31 graphs, we have verified that our results are stable to $\lesssim 0.05$ times the mean values. This shows that, effectively, the GNN outputs are robust under these transformations. On the other hand, it is pertinent to check how sensitive the outputs are to the number of satellites considered. The MW and M31 predictions can increase by up to $\sim 15$\% when the smallest satellites, the SMC and M32 respectively, are removed. If M110 is also removed from the M31 group, there can be an additional $\sim 15$\% increase if the velocities are employed, or a $\sim 5$\% decrease when only positions and stellar masses are considered. This change of tendency if velocities are employed or not reveals that the GNN may rely more on some satellites than others depending on the features considered, as shown in \cite{GNN_CAMELS}. We cannot push this further, as the network is trained on halos with at least one satellite, so it would not be reliable to apply it to only the central galaxy. Analogously, we have also checked the sensitivity of the results when the next satellites in stellar mass are also taken into account. The following satellites in order of decreasing stellar mass are the Sagittarius Dwarf Spheroidal Galaxy (Sgr dSph) and NGC 147 for the MW and M31 respectively, having both stellar masses just below the CAMELS galaxy mass resolution, around $\sim 10^8~ M_\odot$ \cite{2020MNRAS.497.4162V, 2013ApJ...779..102K}. Considering Sgr dSph for the MW halo can lead to a $\sim 15$\% increase, while NGC 147 enhances the M31 mass up to a $\sim 30$\%. Note however that these predictions are not reliable and are only useful for testing purposes, since the GNNs have been trained only with galaxies above the CAMELS mass resolution. Anyway, all these tests show that the presence of satellite galaxies is highly relevant and has a non-negligible impact on the final results.

\begin{table}
	\begin{center}
	%\resizebox{0.5\textwidth}{!}{
	\setlength{\tabcolsep}{10pt}    % row separation
	\renewcommand{\arraystretch}{1.3}   % column separation
		\begin{tabular}{|c|c|c|c|c|}
			\hline
			\multirow{3}{*}{Halo} & \multicolumn{4}{c|}{Mass $[10^{12}\, M_\odot]$} \\
			\cline{2-5}
			 & \multicolumn{2}{c|}{IllustrisTNG} & \multicolumn{2}{c|}{SIMBA} \\ 
			 \cline{2-5}
			 & w/o $v$ & w/ $v$ & w/o $v$ & w/ $v$ \\
			\hline\hline
			MW & $1.0^{+0.9}_{-0.5}$ & $2.6^{+1.2}_{-0.8}$ & $1.3^{+0.9}_{-0.6}$ & $2.3^{+0.9}_{-0.7}$ \\
			\hline
			M31 & $2.3^{+2.4}_{-1.2}$ & $2.3^{+1.2}_{-0.8}$ & $2.5^{+1.8}_{-1.0}$ & $2.2^{+1.3}_{-0.8}$ \\
			\hline
		\end{tabular}
		%}
	\end{center}
	\caption{Constraints on the masses of the MW and M31 halos from models trained in IllustrisTNG and SIMBA simulations, either including or excluding velocities to train the models.}
	\label{table:masspredictions}
\end{table}

\section{Discussion}
\vspace{-0.25cm}

We have applied a novel machine learning model trained on hydrodynamic simulations to predict the masses of the halos hosting the MW and M31 galaxies, obtaining results in good agreement with standard observational methods from rotation curves, Local Group dynamics, kinematic tracers and others techniques. To our knowledge, these are the first constraints on the MW and M31 halo masses from artificial intelligence. The predicted masses do not make any assumptions about the underlying cosmology and astrophysical scenario, since the models have been trained to marginalize over a broad parameter space. This achievement illustrates the suitability of these models to be applied in other galactic systems. Our GNN-based method relies on few observational galactic properties, including the 3D positions, velocities, and stellar masses. When velocities are considered as features, predictions for the MW are slightly larger than those using only positions and stellar mass. This behavior is, however, expected, given the large velocity of LMC with respect to their counterparts in simulations.

Our method presents some caveats which are worth noting. First, we do not include observational errors in the galaxy properties belonging to the halos containing MW and M31. Accounting for these statistical uncertainties would increase the expected confidence level of our predictions. Furthermore, it is not clear which observational features would be more relevant to infer halo masses. It is possible that employing other properties (e.g. neutral hydrogen mass, metallicity, morphology or luminosity), our results may become more precise, robust and reliable. Training the GNNs with a more realistic and tight prior, may also enhance the accuracy; for instance only training on models that match some observable quantity such as the stellar mass function or star-formation rate history. Note that we have implicitly assumed that the CAMELS training set is representative of the MW and M31 galaxies, which could not be the case, since the MW may present atypical features in some regards. Note also that the CAMELS simulations adopt a flat prior over the range of initial conditions and parameters employed, which could give too much weight to some unlikely and extreme cosmological and astrophysical scenarios. Moreover, the effect of nearby halos as interlopers has not been explicitly taken into account, which may impact the predictions, given that the MW and M31 are close and gravitationally bound. One could restrict the set of simulations used for training the models to those with properties similar to the MW and M31 system, either in terms of a their current configuration as a pair of binary galaxies approaching each other, or alternatively having a history that matches their expected evolution. Besides, training on simulations with higher resolution will enable the inclusion of smaller and fainter satellites that can provide additional information and therefore tighten the constraints. This method is hence a proof-of-concept rather than a definitive technique, since there is still room for improvement, which is left for future work.

The implementation of the GNNs underlying this article, {\tt HaloGraphNet}, is available on \href{https://github.com/PabloVD/HaloGraphNet}{GitHub \faicon{github}} at \url{https://github.com/PabloVD/HaloGraphNet} \citep{HaloGraphNet}, making use of PyTorch Geometric \citep{Fey_Fast_Graph_Representation_2019}. Details on the CAMELS simulations can be found in \url{https://www.camel-simulations.org}.

\section*{Acknowledgements}
\vspace{-0.25cm}

We thank Adrian Price-Whelan, Alan W. McConnachie and Sangmo Tony Sohn for their help with Milky Way and Andromeda satellites data. We also thank Nicolas Garavito and Vasily Belokurov for enlightening discussions. The work of PVD is supported by CIDEGENT/2018/019, CPI-21-108. FVN was supported by funding from the WFIRST program through NNG26PJ30C and NNN12AA01C. 
DAA was supported in part by NSF grants AST-2009687 and AST-2108944, and by the Flatiron Institute, which is supported by the Simons Foundation.
The training of the GNNs has been carried out using GPUs from the Tiger cluster at Princeton University.

\bibliography{bibliography}

\end{document}